\documentclass[twocolumn,preprintnumbers,amsmath,amssymb,superscriptaddress,footnote]{revtex4-2}
\usepackage{graphicx} 
\usepackage[dvipdfmx]{epsfig} 
\usepackage{bm}
\usepackage{ulem}
\usepackage{amsmath,color}

\def\m@thcombine#1#2{
  \setbox0=\hbox{$#1$}
  \setbox1=\hbox{$#2$}
  \ifdim\wd0>\wd1
    \setbox0=\hbox to\wd1{\hss\box0\hss}
  \else
    \setbox1=\hbox to\wd0{\hss\box1\hss}
  \fi
  \mathop{\vcenter{
    \offinterlineskip\box0\box1}}}
\def\lesim{\m@thcombine<\sim}
\def\gesim{\m@thcombine>\sim}

\begin{document}
 
\title{Dineutron-dineutron correlation in $^8$He}

\author{Y. Yamaguchi}
\affiliation{Department of Physics, Osaka Metropolitan University, Osaka 558-8585, Japan}

\author{W. Horiuchi}
\email{whoriuchi@omu.ac.jp}
\affiliation{Department of Physics, Osaka Metropolitan University, Osaka 558-8585, Japan}
\affiliation{Nambu Yoichiro Institute of Theoretical and Experimental Physics (NITEP), Osaka Metropolitan University, Osaka 558-8585, Japan}
\affiliation{RIKEN Nishina Center, Wako 351-0198, Japan}
\affiliation{Department of Physics,
    Hokkaido University, Sapporo 060-0810, Japan}

\author{T. Ichikawa}
\affiliation{SIGMAXYZ Inc., 4-1-28 Toranomon, Minato-ku, Tokyo 105-0001, Japan}
\affiliation{Yukawa Institute for Theoretical Physics, Kyoto University,
    Kyoto 606-8502, Japan}

\author{N. Itagaki}
\email{itagaki@omu.ac.jp}
\affiliation{Department of Physics, Osaka Metropolitan University, Osaka 558-8585, Japan}
\affiliation{Nambu Yoichiro Institute of Theoretical and Experimental Physics (NITEP), Osaka Metropolitan University, Osaka 558-8585, Japan}
\affiliation{RIKEN Nishina Center, Wako 351-0198, Japan}

 \preprint{NITEP 175}

 
\begin{abstract}
    \begin{description}

        \item[Background]
            The four-neutron correlation has been attracting much attention for decades.
            In addition to the study on the tetra-neutron system, it is worthwhile to investigate the correlation in bound systems.
        \item[Purpose]
            The $^8$He nucleus is a system where four neutrons are weakly bound around the $^4$He core.
            The dineutron ($2n$) correlation has been long discussed in various weakly-bound neutron-rich nuclei such as $^6$He and $^{11}$Li, whereas the $^8$He nucleus gives us an opportunity to investigate the $2n$-$2n$ type four-neutron correlation.
        \item[Methods]
          We introduce a microscopic $^4$He+$4n$ model
          and describe the ground-state structure of $^8$He.
          The mixing of the two-$2n$ component in the ground state
          is examined. The ground-state wave function is verified
          by investigating various observables including
          high-energy scattering cross sections.
        \item[Results]
            Our model reasonably reproduces the available experimental data,
            the binding energy, charge radius, total reaction cross section,
            and proton-nucleus elastic scattering cross section data.
            We find that the significant mixing of the two-$2n$ cluster configurations around $^4$He in the ground state of $^8$He: The ground state has a squared overlap of about 45\% with a $2n$-$^4$He-$2n$ configuration with the $^4$He-$2n$ distance of 3~fm and opening angle of 80$^\circ$.

        \item[Conclusions]
            The ground state of $^8$He contains a certain amount of the two-$2n$ cluster component, indicating the strong nuclear deformation, which was experimentally observed recently.

    \end{description}
\end{abstract}
\maketitle

The existence of proton deficient nuclear systems has been a subject of debate in nuclear physics. In this context, the tetra-neutron system, comprising four neutrons, has garnered significant interest for decades~\cite{PhysRevC.65.044006, PhysRevLett.116.052501, tetra-nature, PhysRevC.93.044004, PhysRevLett.117.182502, DELTUVA2018238}. A crucial characteristic of the tetra-neutron system is that in the lowest energy configuration, one of the two-neutron pairs, known as dineutron ($2n$), occupies the $s$ state, while the other pair must occupy the $p$ state due to the Pauli exclusion principle. Consequently, the four-neutron correlation may be weakened in a vacuum as the two-$2n$ systems should occupy 
distinctly different orbits with each other.

The two-$2n$ correlations may be enhanced when a core nucleus, such as a $^4$He nucleus, is introduced into the four-neutron system. This is because, in the $^8$He nucleus, four neutrons are bound around the $^4$He core nucleus, and all four can occupy the $p$ state as their lowest energy configuration. As a result, the four-neutron correlation could emerge under a democratic situation, in contrast to the tetra-neutron system in a vacuum. It is important to note that in the $^7$H ($^{3}{\rm H}+4n$) system, the four neutrons are no longer bound, although numerous experimental and theoretical investigations have focused on exploring the resonance states of $^7$H~\cite{PhysRevLett.90.082501, PhysRevLett.99.062502, PhysRevC.103.044313, CAAMANO2022137067, GOLOVKOV2004163, PhysRevC.69.034336, PhysRevC.80.021304, HIYAMA2022137367}.

The $2n$ correlations in the $^8$He nucleus have been theoretically discussed in Refs.~\cite{PhysRevC.76.044323, PhysRevC.77.054317, PhysRevC.78.017306}, even though the neutron number of six in $^7$H and $^8$He corresponds to the subclosure of the $p_{3/2}$ orbits in the nuclear shell model, where the spin-orbit interaction acts attractively. The spin-orbit interaction serves as a driving force for promoting the independent particle motion of each nucleon with good total angular momentum~\cite{PhysRevC.83.014302} and leads to the breaking of the spin-singlet $2n$ clusters. However, if the four valence neutrons in the $p$ shell have extended spatial distributions outside the interaction region of the spin-orbit force, it would be possible to form the $2n$ clusters there. Recent experimental results support such spatially extended four-neutron wave functions: The four neutrons in the $^8$He nucleus are bound only by 3.11~MeV from the $^4{\rm He}+4n$ threshold energy~\cite{TILLEY2004155}, and the matter radius is evaluated as 2.53(2) fm with a neutron skin thickness of 0.82(2) fm, implying an extended four-neutron distribution~\cite{WAKASA2022105329}.

We note that the $2n$-$2n$ correlation has not been investigated sufficiently, although the $2n$ correlations are extremely important in weakly bound systems and have been widely discussed in neutron-rich nuclei~\cite{BERTSCH1991327, PhysRevC.71.064326, PhysRevLett.124.212503}. In Refs.~\cite{PhysRevC.76.044323,PhysRevC.78.017306}, the squared overlap between the ground-state wave function of $^8$He and two $2n$ cluster configurations was examined. In those studies, two $2n$ clusters were isotopically distributed around $^8$He by means of the Tohsaki-Horiuchi-Schuck-Röpke (THSR) wave function~\cite{PhysRevLett.87.192501}. A similar approach was adopted in Ref.~\cite{PhysRevC.88.034321}. Several {\it ab initio} calculations were performed for the $^8$He nucleus.
Many physical properties can be interpreted in connection with the realistic
interactions but unfortunately the $2n$-$2n$ correlation was not discussed there~\cite{PhysRevC.64.014001,PhysRevC.73.021302,PhysRevC.79.014308,PhysRevC.105.034313}. Ref.~\cite{PhysRevC.77.054317} conducted the Hartree-Fock-Bogoliubov calculations with the Wood-Saxon potential, where the four-body density of the valence neutrons and the $2n$-$2n$ correlation were discussed but that obtained nuclear radius is unrealistically larger than the experimental value. 
An investigation using a realistic wave function is needed to extract an amount of the $2n$-$2n$
  correlation quantitatively.

  The recent experiment posited that the ground state of $^8$He is strongly deformed~\cite{HOLL2021136710}. This suggests that the $^{8}$He configuration is not a simple subclosure configuration of $p_{3/2}$. Instead, it may exhibit enhanced four-neutron correlations.  Moreover, the possibility of the so-called soft dipole mode in $^8$He was recently pointed out in Ref.~\cite{10.1093/ptep/ptac130}, implying the formation of a four-neutron cluster around the $^4$He core.

The purpose of this paper is to investigate the $2n$-$2n$ correlation in the $^8$He nucleus. We perform microscopic cluster model calculations that explicitly include various geometric configurations of two-$2n$ clusters. We remark that a strong dineutron correlation
    was implied within a $^{4}$He core plus $4n$ picture~\cite{PhysRevC.98.061302}. We take special care to describe the valence four-neutron wave functions around the $^4$He core by superposing numerous configurations. To verify our wave function, we calculate the proton-$^8$He elastic scattering and total reaction cross sections 
and compare them with available experimental data. Finally, we discuss whether the four neutrons occupying the $p$ states can favorably form two $2n$ clusters under a democratic condition and the relationship between the two-$2n$ correlations and nuclear deformation.
 
Here, we introduce a microscopic $^4$He+$4n$ model, wherein the wave function is fully antisymmetrized. We construct basis states with different neutron configurations and superpose them based on the generator coordinate method (GCM) after the angular momentum projection~\cite{Brink}. The coefficients of each basis state are determined by diagonalizing the norm and Hamiltonian matrices. Using the obtained ground-state wave function of $^{8}$He, we calculate the one-body density distributions for the cross-section calculations with the Glauber model~\cite{Glauber}.

The Hamiltonian $(H)$ consists of kinetic energy $(t_i)$ and
potential energy ($v_{ij}$) terms as
\begin{equation}
    H = \sum_{i=1}^{8}t_i- T_{\rm c.m.} + \sum_{i < j} v_{ij},
\end{equation}
where the center-of-mass kinetic energy $T_{\rm c.m.}$ is subtracted to guarantee the translation-invariance of the wave functions. For the potential part, the interaction consists of the central, spin-orbit, and Coulomb terms. We employ the Volkov No.2~\cite{VOLKOV196533} effective nucleon-nucleon interaction with the Majorana exchange parameter of $M = 0.6$, which is known to reproduce the low-energy scattering phase shift of $^4{\rm He}+^4$He~\cite{PTP.61.1049}. The original Volkov interaction does not have the Heisenberg and Bartlett exchange terms, but we introduce $B=H=0.07$ just as in Ref.~\cite{PhysRevC.80.021304} to remove the spurious bound state of the two-neutron system. For the spin-orbit part, we use the spin-orbit term of the G3RS interaction~\cite{PTP.39.91}, which is a realistic interaction originally developed to reproduce the nucleon-nucleon scattering phase shifts. The strength of the spin-orbit interaction is set to 2000 MeV, reproducing the low-energy scattering phase shift of $p+^4$He~\cite{PTP.39.91}.

The wave function of $^{8}$He is described by a superposition of fully antisymmetrized ($\mathcal{A}$) $^{4}{\rm He}(\alpha)+4n$ wave function as
\begin{align}
    \Phi=\mathcal{A}\left\{\Phi_\alpha (\bm{R}_\alpha)
    \Phi_{4n}(\bm{R}_1,\bm{R}_2,\bm{R}_3,\bm{R}_4)\right\},
\end{align}
where the $\alpha$ and $4n$ wave functions, $\Phi_\alpha$ and $\Phi_{4n}$, are expressed by the product of the single-particle wave function with a Gaussian form as used in many other cluster models, including the Brink model~\cite{Brink}
\begin{equation}
    \phi^{\tau, \sigma} \left( \bm{r} \right)
    =
    \left(  \frac{2\nu}{\pi} \right)^{\frac{3}{4}}
    \exp \left[- \nu \left(\bm{r} - \bm{R} \right)^{2} \right] \chi^{\tau,\sigma},
    \label{spwf}
\end{equation}
where $\bm{R}$ is a Gaussian center parameter related to the expectation value for the position of the nucleon, and $\chi^{\tau,\sigma}$ are the spin and isospin parts of the wave function. For the size parameter $\nu$, we here use $\nu =0.255$ fm$^{-2}$, slightly smaller than the value for the free $^4$He nucleus, which is reasonable, as the $^{4}$He core swells due to the interaction and Pauli principle from the valence neutrons~\cite{PhysRevC.59.1432,PhysRevC.89.064303}. The $\alpha$ cluster as the core nucleus can be expressed by four nucleons with the spin and isospin saturated configuration sharing the same $\bm{R}_\alpha$ value. Similarly, the $2n$ cluster can also be expressed by two neutrons with spin-up and spin-down with $\bm{D}_1\equiv \bm{R}_1=\bm{R}_2$ and $\bm{D}_2\equiv\bm{R}_3=\bm{R}_4$.

Each Slater determinant is projected to the eigenstates of parity and angular momentum by using the projection operator
\begin{equation}
    P_{J^\pi M}^K
    =
    P^\pi \frac{2J+1}{8\pi^2}
    \int d\Omega \, \left\{D_{MK}^{J}\right\}^* R \left(\Omega \right).
\end{equation}
Here ${D_{MK}^J}$ is the Wigner $D$-function and $R\left(\Omega\right)$ is the rotation operator acting on the spatial and spin parts of the wave function. This integration over the Euler angle $\Omega$ is numerically performed. The operator $P^\pi$ is for the parity projection. Here we take $P^+ = \left(1+P^r\right) / \sqrt{2}$ for the positive-parity states, where $P^r$ is the parity-inversion operator.
\par
The generated many different Slater determinants ($\{P_{J^\pi M}^K \left| \Phi_i\right> \}$) are superposed based on GCM \cite{Brink}. After normalizing each basis state, the total wave function $\Psi_{J^\pi M}$ is written as
\begin{equation}
    \Psi_{J^\pi M} = \sum_i c_i P_{J^\pi M}^K \left| \Phi_i\right>
    \label{GCM}
\end{equation}
The coefficients $\left\{ c_i \right\}$ for a linear combination of the Slater determinants are obtained together with the energy eigenvalue $E$ when we diagonalize the norm and Hamiltonian matrices, namely by solving the Hill-Wheeler equation~\cite{PhysRev.89.1102}
\begin{align}
     & \sum_{j} \left[\left<\Phi_i \right| (P_{J^\pi M}^{K})^\dagger H P_{J^\pi M}^K \left| \Phi_j\right>\right.\notag \\
     & \left.- E \left<\Phi_i\right| (P_{J^\pi M}^{K})^\dagger P_{J^\pi M}^K \left| \Phi_j\right>\right] c_j = 0.
    \label{HW.eq}
\end{align}

In this study, we prepare $^4{\rm He}+2n+2n$ wave functions with various isosceles triangle configurations and randomly generated $^4{\rm He}+4n$ (the $\alpha$ core + four free neutrons) wave functions.

\begin{figure}[tb]
    \centering
    \includegraphics[keepaspectratio,width=\linewidth]{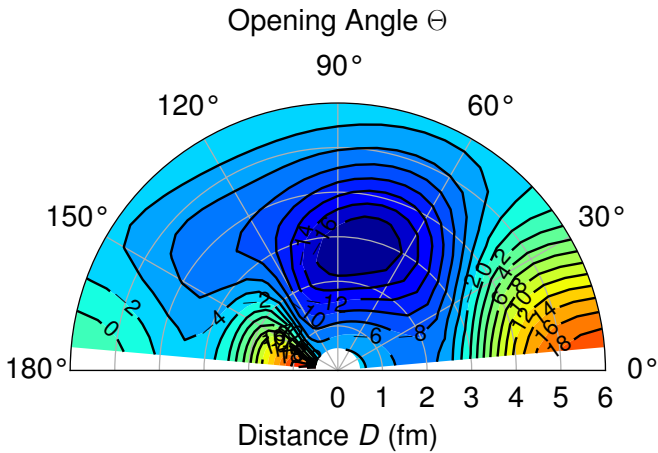}
    \caption{
      Potential energy surface of $J^{\pi}=0^+$ for the $^4{\rm He}+2n+2n$
      configuration as a function of the distance $D$ 
      and the opening angle $\Theta$ between two $2n$s 
              Contours are drawn by 2 MeV intervals.
    }
    \label{he8PES}
\end{figure}

The geometric configurations of the $^4{\rm He}+2n+2n$ wave functions are the isosceles triangles where the two sides of the $^4$He-$2n$ distance $D\equiv |\bm{D_}1-\bm{R}_\alpha|=|\bm{D}_2-\bm{R}_\alpha|$ are taken as $D=1$, 2, 3, 4, 5, and 6 fm, and its opening angle $\Theta$ is taken as $\Theta=30^\circ$, 60$^\circ$, $90^\circ$, $120^\circ$, and $150^\circ$ with the spin saturated configuration. Figure~\ref{he8PES} shows the calculated potential energy surface of $J^{\pi}=0^+$ for the $^4{\rm He}+2n+2n$ configuration as a function of $D$ and $\Theta$. We can see a local energy minimum of $E=-17.97$~MeV at around $D=3.0$~fm and $\Theta=80^\circ$. Note that the rms distance of the two neutrons
  in the $2n$ cluster is 2.43 fm, which implies developed $2n$ clusters
in the surface region in the ground state of $^{8}$He.
  Large positive values of the potential energy surface
come from the Pauli principle.

For the $4n$ configurations, we randomly generate the Gaussian center parameters $\bm{R}_1, \bm{R}_2, \bm{R}_3$, and $\bm{R}_4$ of the four neutrons, where two of them are spin-up and the remaining two are spin-down. To accelerate the energy convergence, these random numbers are generated by following an exponential distribution having a width of 1.4~fm for the distance between neutrons with the up and down spins and the distance between their center position and the $\alpha$ core. The center of mass of each Slater determinant is shifted to the origin before the superposition.

\begin{figure}[tb]
    \centering
    \includegraphics[width=\linewidth]{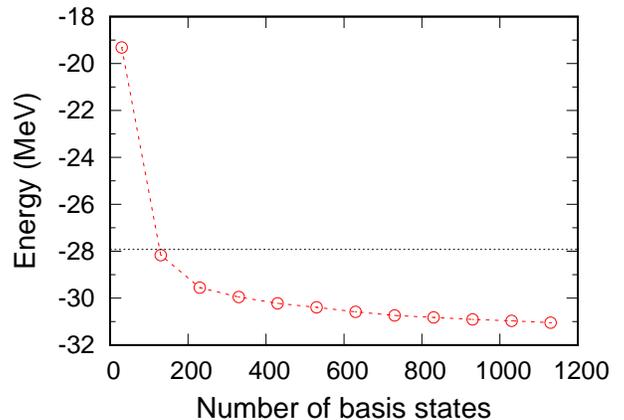}
    \caption{
        Energy convergence of the lowest $J^\pi=0^+$ state of
        $^8$He as a function of the number of basis states superposed.
        The horizontal thin dotted line indicates
        the theoretical $^4$He+$4n$ threshold energy, $-27.92$~MeV.
    }
    \label{he8conv}
\end{figure}

Before discussing the ground state of $^8$He,
  we calculate $^{6}$He by
  randomly generated two-neutron configurations around an $\alpha$ cluster
  to verify our choice of the model Hamiltonian.
  The binding energy is found to be $-1.33$ MeV from the $\alpha+n+n$ threshold
  with 200 basis states. The value is slightly overbinding compared with
  the experimental data of $-0.975$~MeV~\cite{Wang_2021} 
  but is within the acceptable range because the calculated
  total reaction cross section on a carbon target
  is 719 mb at 800 MeV/nucleon, which is in good agreement
  with the experimental interaction cross section
  at 790 MeV/nucleon, 722$\pm 6$~mb~\cite{TANIHATA1985380}.
  The details of the cross-section calculation
  will be described for a $^{8}$He case later.
  The root-mean-square (rms) matter radius
  is calculated as 2.44 fm, which is slightly smaller than
  the empirical value 2.48$\pm$0.03 fm~\cite{TANIHATA1988592}. 

Figure~\ref{he8conv} draws the energy convergence of the $0^+$ state of $^8$He
as a function of the number of basis states superposed.
The first 30 basis states correspond to the $^4{\rm He}+2n+2n$ basis states, and the subsequent 1,100 basis states represent the $^4$He+$4n$ basis states. To achieve convergence, more than 1,000 basis states are required.

The calculated energy of the $^8$He nucleus is found to be $-3.14$ MeV from the threshold, which is in reasonable agreement with experimental data, $-3.10$~MeV~\cite{Wang_2021}. This result demonstrates that the wave function used in this study provides a good description of the four-valence neutrons of the $^8$He nucleus.

\begin{figure}[tb]
    \centering
    \includegraphics[width=\linewidth]{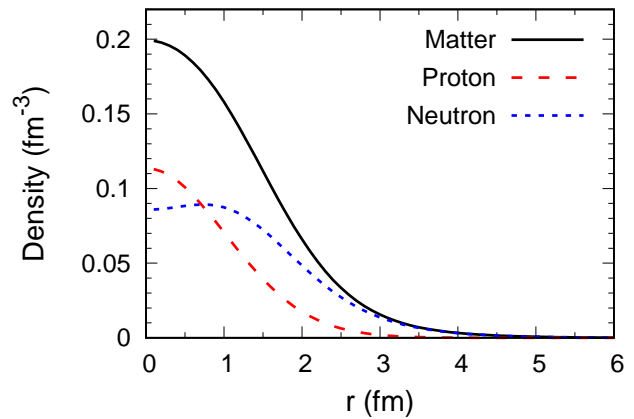}
    \caption{
      Point-matter, proton, and neutron one-body density distributions of $^8$He.
    }
    \label{density}
\end{figure}

Figure~\ref{density} displays the point-matter, proton, and neutron
density distribution of $^8$He. We see
more extended neutron density distribution than proton one,
indicating a thick neutron skin.
The rms point-proton and matter radii of
these density distributions are calculated
as 1.82~fm and 2.50~fm, respectively.
The calculated point-proton radius is in good agreement with the experimental ones deduced from the charge radius, 1.80(3)~fm~\cite{ANGELI201369} and 1.81(3) fm~\cite{PhysRevLett.99.252501}.

To further verify our wave function, we calculate the total interaction cross sections on a carbon target as well as proton-nucleus elastic scattering cross sections.
These calculations are performed based on the Glauber model~\cite{Glauber}
with the optical limit approximation~\cite{Glauber,Suzuki03}.
The nucleon-target formalism~\cite{PhysRevC.61.051601} is employed
for the total reaction cross section calculation on a carbon target.
The inputs to the theory are the one-body density distributions and the profile function which describes the nucleon-nucleon scattering properties.
The profile function~\cite{PhysRevC.77.034607} has been well tested as shown in many examples of proton-nucleus~\cite{JPSJ.78.044201,PhysRevC.93.044611,Hatakeyama19}
and nucleus-nucleus reactions~\cite{PhysRevC.74.034311,PhysRevC.75.044607,JPSJ.78.044201,PhysRevC.81.024606,PhysRevC.86.024614,JPSCP.6.030079,PhysRevC.97.054614}. For more details, see, for example, Refs.~\cite{PhysRevC.106.044330,PhysRevC.107.L021304}  and references therein, showing the most recent application of this model. The essential input for the cross-section calculations,
the one-body density distribution
of the ground state of $^8$He,  is obtained after solving the Hill-Wheeler equation (Eq.~(\ref{HW.eq})). The density distribution is calculated in the angular momentum projected space, which is free from the center of mass motion,
and the detail is described in Ref.~\cite{BAYE1994624}.

Using these density distributions, the total reaction cross section on a carbon target at an incident energy of 800~MeV/nucleon is calculated as 798~mb. This value is slightly smaller than the experimental cross section of 817(6)~mb at 790~MeV/nucleon~\cite{TANIHATA1985380} but still within the acceptable range considering the uncertainties of previous observations~\cite{PhysRevLett.99.252501}.

\begin{figure}[tb]
    \centering
      \includegraphics[width=\linewidth]{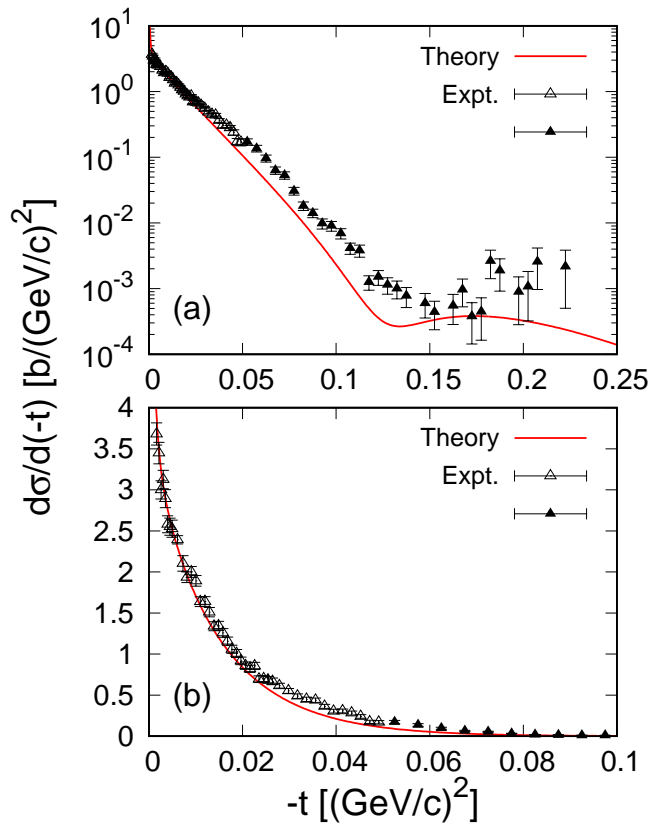} 
    \caption{
        Proton-nucleus
        differential elastic scattering cross sections of $^{8}$He
        at the incident energy of 680 MeV
        as a function of  the four momentum transfer squared
        plotted in (a) logarithmic and (b) linear scales.
        The experimental data are taken from Refs~\cite{NEUMAIER2002247,KISELEV201172}.
    }
    \label{dcs}
\end{figure}

Figure~\ref{dcs} 
compares the calculated differential cross section for proton-$^8$He scattering at an incident proton energy of 680~MeV with available experimental data at low~\cite{NEUMAIER2002247} and high~\cite{KISELEV201172}
four momentum transfer regions. Although there are some deviations at intermediate momentum transfer regions, the overall agreement between theory and experiment is achieved. 
To see it more quantitatively at low four momentum transfer regions,
in Fig.~\ref{dcs} (b), we also plot the cross sections in a linear scale.
This good agreement confirms the validity of the wave function used in this study.

\begin{figure}[tb]
    \centering
    \includegraphics[width=\linewidth]{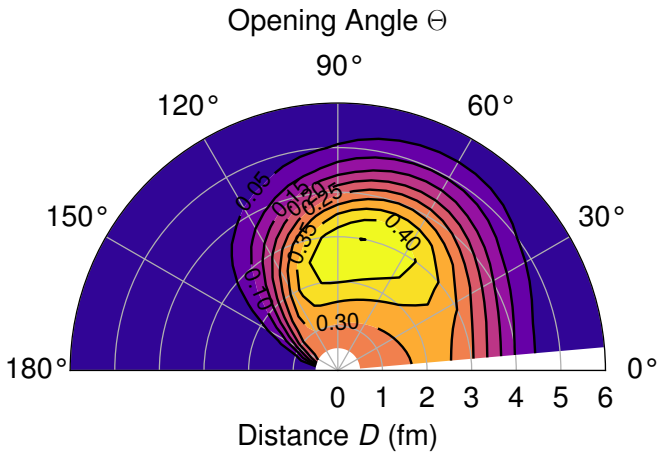}
    \caption{
      Squared overlap between the ground state
      of $^8$He and $^4{\rm He}+2n+2n$ cluster configuration
      with isosceles triangular configurations.
      as a function of the $^4$He-$2n$ distance $D$ 
  and the opening angle $\Theta$ between two $2n$s
  of the cluster configuration.
   Contours are drawn by 0.05 intervals.
    }
    \label{di-di-comp}
\end{figure}

Using the realistic wave function obtained in this way, the two-$2n$ correlation in $^8$He is investigated by analyzing the wave function components. The squared overlap between the ground state of $^8$He and the $^4$He+$2n$+$2n$ cluster configuration with isosceles triangular configurations 
is evaluated.
Figure~\ref{di-di-comp} 
displays the squared overlap results, showing that a $^4$He-$2n$ distance of $\sim 3$~fm gives the largest squared overlap and a peak structure is found around the $2n$-$^4$He-$2n$ opening angle of $\Theta=80^\circ$. This large overlap area corresponds to a deep pocket in the potential energy surface coming from the triangular cluster configuration (Fig.~\ref{he8PES}).
\par
The value of squared overlap at the peak position is about 0.45, indicating that $^8$He contains a significant amount of the two-$2n$ component with a large $^4$He-$2n$ distance. Despite the neutron number 6 corresponding to the closure of $p_{3/2}$ in the $jj$-coupling shell model, the spin-orbit interaction does not completely break the $2n$ clusters due to the weakly bound nature of the system; neutrons also stay beyond the interaction range of the spin-orbit interaction, $\sim 2.5$~fm~\cite{PTP.39.91}. The rapid drop of the squared overlap beyond $\Theta=90^\circ$ 
indicates that obtuse triangular shapes
are not favored, which also suggests the correlation between the two $2n$ clusters. We also compute the squared overlap between
  the $0\hbar\omega$ harmonic-oscillator configuration [$(0s_{1/2})^4(0p_{3/2})^4$] and the ground state wave function. This overlap value is 0.40, which is comparable to the two-$2n$ component. We remark that the coupling with the continuum
  states significantly enhances the component of
  the continuum $p_{3/2}$ orbits~\cite{PhysRevC.85.034338}.

  To relate the two-$2n$ component with the nuclear shape,
  it is instructive to quantify the degree of deformation for each 
  basis.
The $\beta$ parameters is given by ~\cite{Bohr75}
\begin{align}
    \beta&=\sqrt{a_0^2+2a_2^2}, 
\end{align}
where the dimensionless deformation parameter is defined by
\begin{equation}
    a_m=\frac{4\pi}{5}\frac{Q_{2m}}{AR^2}
    \label{defa}
\end{equation}
with $R = 1.2 A^{1/3}$~fm and
the quadrupole deformation parameters
\begin{align}
    Q_{20} & =\sqrt{\frac{5}{16\pi}}(2\langle z^2\rangle-\langle x^2\rangle-\langle y^2\rangle), \\
    Q_{22} & =\sqrt{\frac{15}{32\pi}}(\langle y^2\rangle-\langle x^2\rangle).
\end{align}

\begin{figure}[tb]
    \centering
      \includegraphics[width=\linewidth]{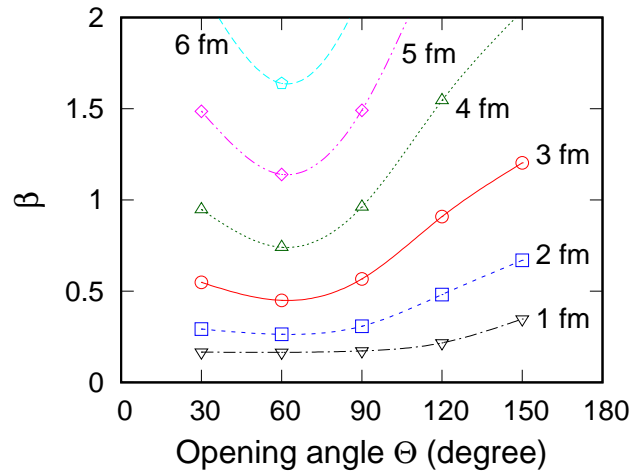} 
    \caption{
      Quadrupole deformation parameter $\beta$ of
      the $^4$He+$2n$+$2n$ cluster configuration
      with isosceles triangular configurations
      as a function of
      the opening angle $\Theta$ of $2n$-$^4$He-$2n$.
    }
    \label{beta}
\end{figure}
   
Figure~\ref{beta} displays the calculated $\beta$ of the $^4$He+2$n$+2$n$ configuration, showing that the value rapidly increases with increasing $^4$He-$2n$ distance. The mixing of such components could explain the large deformation of $^8$He discussed in the literature ($\beta = 0.4$~\cite{HOLL2021136710}). The large experimental $\beta$ value is consistent with the large distance between $^4$He and $2n$s; however, it is not necessarily direct evidence of $2n$-$2n$ correlation with smaller opening angles. As shown in Fig.~\ref{beta},
large opening angles can also give large $\beta$ values.

    To conclude, the four-neutron correlation has been a disputable topic in nuclear physics for decades. This study has investigated for the first time dineutron-dineutron ($2n$-$2n$) correlations in $^8$He using a reliable wave function obtained from a microscopic $^{4}{\rm He}+4n$ cluster model. To construct a $^8$He wave function, explicit two-$2n$ configurations were superposed along with many $4n$ configurations. The agreement between the theory and experiment is satisfactory.
Overall, this study provides insights into the two-$2n$ correlation
in $^8$He and its effects on the nuclear structure. 
The results show that a significant amount of the two-$2n$ components in the ground state of $^8$He could explain the strong nuclear deformation observed in recent studies. However, it should be noted that the strong deformation is not necessarily a direct evidence of $2n$-$2n$ correlations with small opening angles, as it can also be interpreted as configurations locating $2n$s at large distances.

Further experimental studies are needed to clarify the existence of the two-$2n$ correlations. Exploring the possibility of such $2n$-$2n$ correlations in other neutron-rich nuclei near the neutron dripline is also an interesting topic for the universal understanding of the emergence of the $2n$-$2n$ correlation.\hfill\break

This work was in part supported by JSPS KAKENHI Grants
    Nos.\ 18K03635, 22H01214, and 22K03618.
    The numerical calculations were performed using the computer facility of
    Yukawa Institute for Theoretical Physics,
    Kyoto University (Yukawa-21).

\bibliography{he8.bib}

\end{document}